\documentclass[prb,twocolumn,showpacs,preprintnumbers,amsmath,aps]{revtex4}
\usepackage{graphicx,hyperref}% Include figure files
\usepackage{bm}% bold math
\usepackage{subfigure}% bold math

\def\nbC{{\mathchoice {\setbox0=\hbox{$\displaystyle\rm C$}%
\hbox{\hbox to0pt{\kern0.4\wd0\vrule height0.9\ht0\hss}\box0}}
{\setbox0=\hbox{$\textstyle\rm C$}\hbox{\hbox
to0pt{\kern0.4\wd0\vrule height0.9\ht0\hss}\box0}}
{\setbox0=\hbox{$\scriptstyle\rm C$}\hbox{\hbox
to0pt{\kern0.4\wd0\vrule height0.9\ht0\hss}\box0}}
{\setbox0=\hbox{$\scriptscriptstyle\rm C$}\hbox{\hbox
to0pt{\kern0.4\wd0\vrule height0.9\ht0\hss}\box0}}}}
\def\nbQ{{\mathchoice {\setbox0=\hbox{$\displaystyle\rm
Q$}\hbox{\raise
0.15\ht0\hbox to0pt{\kern0.4\wd0\vrule height0.8\ht0\hss}\box0}}
{\setbox0=\hbox{$\textstyle\rm Q$}\hbox{\raise
0.15\ht0\hbox to0pt{\kern0.4\wd0\vrule height0.8\ht0\hss}\box0}}
{\setbox0=\hbox{$\scriptstyle\rm Q$}\hbox{\raise
0.15\ht0\hbox to0pt{\kern0.4\wd0\vrule height0.7\ht0\hss}\box0}}
{\setbox0=\hbox{$\scriptscriptstyle\rm Q$}\hbox{\raise
0.15\ht0\hbox to0pt{\kern0.4\wd0\vrule height0.7\ht0\hss}\box0}}}}
\def\nbT{{\mathchoice {\setbox0=\hbox{$\displaystyle\rm
T$}\hbox{\hbox to0pt{\kern0.3\wd0\vrule height0.9\ht0\hss}\box0}}
{\setbox0=\hbox{$\textstyle\rm T$}\hbox{\hbox
to0pt{\kern0.3\wd0\vrule height0.9\ht0\hss}\box0}}
{\setbox0=\hbox{$\scriptstyle\rm T$}\hbox{\hbox
to0pt{\kern0.3\wd0\vrule height0.9\ht0\hss}\box0}}
{\setbox0=\hbox{$\scriptscriptstyle\rm T$}\hbox{\hbox
to0pt{\kern0.3\wd0\vrule height0.9\ht0\hss}\box0}}}}
\def\nbS{{\mathchoice
{\setbox0=\hbox{$\displaystyle     \rm S$}\hbox{\raise0.5\ht0%
\hbox to0pt{\kern0.35\wd0\vrule height0.45\ht0\hss}\hbox
to0pt{\kern0.55\wd0\vrule height0.5\ht0\hss}\box0}}
{\setbox0=\hbox{$\textstyle        \rm S$}\hbox{\raise0.5\ht0%
\hbox to0pt{\kern0.35\wd0\vrule height0.45\ht0\hss}\hbox
to0pt{\kern0.55\wd0\vrule height0.5\ht0\hss}\box0}}
{\setbox0=\hbox{$\scriptstyle      \rm S$}\hbox{\raise0.5\ht0%
\hboxto0pt{\kern0.35\wd0\vrule height0.45\ht0\hss}\raise0.05\ht0%
\hbox to0pt{\kern0.5\wd0\vrule height0.45\ht0\hss}\box0}}
{\setbox0=\hbox{$\scriptscriptstyle\rm S$}\hbox{\raise0.5\ht0%
\hboxto0pt{\kern0.4\wd0\vrule height0.45\ht0\hss}\raise0.05\ht0%
\hbox to0pt{\kern0.55\wd0\vrule height0.45\ht0\hss}\box0}}}}
\def\nbZ{{\mathchoice {\hbox{$\sf\textstyle Z\kern-0.4em Z$}}
{\hbox{$\sf\textstyle Z\kern-0.4em Z$}}
{\hbox{$\sf\scriptstyle Z\kern-0.3em Z$}}
{\hbox{$\sf\scriptscriptstyle Z\kern-0.2em Z$}}}}
\begin{document}

\title{Activated dynamic scaling in the random-field Ising model: a nonperturbative functional renormalization group approach}

\author{Ivan Balog} \email{balog@ifs.hr}
\affiliation{Institute of Physics, P.O.Box 304, Bijeni\v{c}ka cesta 46, HR-10001 Zagreb, Croatia}
\affiliation{LPTMC, CNRS-UMR 7600, Universit\'e Pierre et Marie Curie,
bo\^ite 121, 4 Pl. Jussieu, 75252 Paris c\'edex 05, France}

\author{Gilles Tarjus} \email{tarjus@lptmc.jussieu.fr}
\affiliation{LPTMC, CNRS-UMR 7600, Universit\'e Pierre et Marie Curie,
bo\^ite 121, 4 Pl. Jussieu, 75252 Paris c\'edex 05, France}

\date{\today}

\begin{abstract}

The random-field Ising model shows extreme critical slowdown that has been described by activated dynamic scaling: the characteristic time for the relaxation to equilibrium diverges exponentially with the correlation length, $\ln \tau\sim \xi^\psi/T$ , with $\psi$ an \textit{a priori} unknown barrier exponent. Through a nonperturbative functional renormalization group, we show that  for spatial dimensions $d$  less than  a critical value $d_{DR} \simeq 5.1$, also associated with dimensional-reduction breakdown, $\psi=\theta$ with $\theta$ the temperature exponent near the zero-temperature fixed point that controls the critical behavior. For $d>d_{DR}$ on the other hand, $\psi=\theta-2\lambda$ where $\theta=2$ and $\lambda>0$ a new exponent. At the upper critical dimension $d=6$, $\lambda=1$ so that $\psi=0$, and activated scaling gives way to conventional scaling. We give a physical interpretation of the results in terms of collective events in real space, avalanches and droplets. We also propose a way to check the two regimes by computer simulations of long-range 1-$d$ systems.

\end{abstract}

\pacs{11.10.Hi, 75.40.Cx}

\maketitle

\section{introduction}

Activated dynamic scaling\cite{fisher-review} is a phenomenological description of the extreme slowdown of dynamics observed in some disordered or glassy systems:  systems in the presence of a quenched random field,\cite{villain-fisher,RFIM} spin glasses in their ordered phase,\cite{spinglass} pinned elastic manifolds,\cite{chauve_creep,gorokhov_creep,balents-ledoussal} and possibly supercooled liquids as they approach their glass transition.\cite{glass} According to this scaling, the dynamics involve thermal activation over barriers that grow with the typical length scale $\ell$, leading to a characteristic time at the scale $\ell$ behaving as $\ln \tau_{\ell} \sim (E/T)\ell^{\psi}$ with $\psi$ an \textit{a priori} unknown barrier exponent. Activated scaling leads to a broad distribution of  relaxation times, which shows up in the time or frequency dependence of the response and correlation functions, and has also consequences for the nonequilibrium dynamics.\cite{fisher-review,chauve_creep,gorokhov_creep}

The random-field Ising model (RFIM)\cite{nattermann98} is one system whose dramatic critical slowing down is expected to be described by activated dynamic scaling.\cite{fisher-review,villain-fisher} Its critical point is controlled, in the renormalization group sense, by a zero-temperature fixed point at which the ``dangerously irrelevant'' renormalized temperature is characterized by an exponent $\theta>0$. The dangerous irrelevancy leads to a breakdown of the hyperscaling relation between critical exponents and to anomalous thermal fluctuations, all controlled by the exponent $\theta$ and further rationalized at a physical level by the ``droplet scenario''.\cite{fisher-review,villain-fisher,spinglass} The simplest droplet assumption would be to set $\psi=\theta$. Actually, this equality has been found in the dynamics of a simpler disordered system, an elastic manifold pinned in a random potential,\cite{drossel,balents-ledoussal} but in the case of the RFIM there has been no attempt to compute the barrier exponent $\psi$.

The functional renormalization group (FRG) is a tool of choice to provide a theoretical treatment beyond phenomenology and compute the barrier exponent $\psi$. In its perturbative form, it has been successfully applied to the dynamics  of the pinned elastic manifolds.\cite{chauve_creep,balents-ledoussal} For the RFIM, as was shown for the (static) equilibrium behavior, the FRG must be nonperturbative.\cite{tarjus04,tissier06,tissier11} In this work we therefore generalize the nonperturbative FRG (NP-FRG) approach to describe the critical slowing down of the RFIM.

We find that the critical slowing down of the RFIM is indeed of activated type with two different regimes as a function of spatial dimension $d$. For $d$  less than  a critical value $d_{DR} \simeq 5.1$, also associated with the breakdown of the $d\rightarrow d-2$ dimensional-reduction property,\cite{footnote_defDR,tarjus04,tissier06,tissier11} the barrier exponent coincides with the temperature exponent, $\psi=\theta$, as in elastic manifolds pinned in a random potential (see above). On the other hand, for $d>d_{DR}$, $\psi=\theta-2\lambda$, where $\theta=2$ (the dimensional-reduction value) and $\lambda>0$ is a new exponent that is computed within the NP-FRG. At the upper critical dimension $d=6$, one finds $\lambda=1$ around the Gaussian fixed point, so that $\psi=0$, and activated scaling gives way to conventional scaling. We stress that in the range $6>d>d_{DR}$ where the main critical exponents describing the static behavior coincide with the dimensional-reduction predictions, the critical dynamics is nonetheless activated and that this feature is completely missed by perturbation theory which instead predicts conventional dynamic scaling, $\tau\sim \xi^z$ with $z\simeq2+2\eta$.\cite{krey84,footnote_DR}

\section{model and dynamical field theory}

As we are interested in the long-time, collective behavior of the RFIM, a coarse-grained field theory provides an appropriate starting point. The relaxation dynamics of the scalar field $\varphi_{xt}$ is thus described by a Langevin equation (for simplicity we consider the case of a nonconserved order parameter, known as model A\cite{hohenberg-halperin})
\begin{equation}
\label{eq_langevin}
\partial_t\varphi_{xt}= -\Omega_B \frac{\delta S[\varphi]}{\delta\varphi_{xt}}+\eta_{xt},
\end{equation}
where $\eta_{xt}$ is a Gaussian random noise term with zero mean and variance $\langle\eta_{xt}\eta_{x't'}\rangle=2T\Omega_B \delta^{(d)}(x-x')\delta(t-t')$. The "action'' or effective Hamiltonian $S[\varphi]$ is given by
\begin{equation}
\begin{aligned}
\label{eq_ham_dis}
&S[\varphi;h+J]=  S_B[\varphi]-\int_{x} [h(x)+J_x] \varphi_x \,, \\&
S_B[\varphi]= \int_{x}\bigg\{\frac{1}{2}[\partial_x \varphi_x]^2+ \frac{r}{2} \varphi_x^2 + \frac{u}{4!} \varphi_x^4 \bigg\},
\end{aligned}
\end{equation}
where $ \int_{x} \equiv \int d^d x$, $J_x$ is an external source, $h_x$ is a random ``source'' (a random magnetic field) taken with a Gaussian distribution characterized by a zero mean and a variance $\overline{h_x h_{x'}}= \Delta_B \delta^{(d)}( x-x')$.

The generating functional of the multi-point and multi-time correlation and response functions can be built as usual by following the MSR formalism.\cite{MSR,janssen-dedom} After introducing an auxiliary ``response'' field $\hat \varphi_{xt}$ and taking into account the fact that the solution of Eq. (\ref{eq_langevin}) is unique,\cite{zinnjustin89} one obtains the ``partition function''
\begin{equation}
\begin{aligned}
\label{eq_part_dis2}
\mathcal Z_{h,\eta}[\hat{J},J]=&\int \mathcal{D}\varphi\mathcal{D}\hat{\varphi} \exp\big\{-\int_{xt}\hat{\varphi}_{xt}\Big[ \partial_t\varphi_{xt}+\Omega_B \frac{\delta S_B[\varphi]}{\delta\varphi_{xt}}  \\& 
-\eta_{xt} -h_x\Big]+\int_{xt}(\hat{J}_{xt}\varphi_{xt}+J_{xt}\hat{\varphi}_{xt})\Big \}\,
\end{aligned}
\end{equation}
where we have used the It\={o} prescription (which amounts to setting to 1 the Jacobian of the transformation between the thermal noise and the field).\cite{zinnjustin89} 

The conventional route for studying the dynamics of disordered systems is then to average the partition function in Eq. (\ref{eq_part_dis2}) over both the thermal noise and the disorder and to take advantage of the property that $\mathcal Z_{h,\eta}[\hat{J}=0,J]=1$.\cite{dedom78} However, in previous NP-FRG work on the RFIM,\cite{tarjus04,tissier06,tissier11} it was shown that the key point for taking relevant events such as avalanches and droplets into account is to describe the full functional dependence of the cumulants of the renormalized disorder, a point that is overlooked by the standard replica, superfield or dynamic formalisms. The most convenient procedure to obtain this full functional dependence is to introduce copies or replicas of the system: the copies have the same disorder $h$ but are coupled to \textit{distinct} sources, in contrast with the usual replica trick.\cite{tarjus04,tissier11} We therefore combine  dynamics \textit{and} replicas/copies. The latter are now characterized not only by distinct sources but also by independent thermal noises.\cite{footnote_replicas}

After averaging over the thermal noises and the disorder, one obtains
\begin{equation}
\begin{aligned}
\label{eq_part_aver}
&Z[{\hat{J}_a,J_a}]= \\& \int \prod_a \mathcal{D}\varphi_a\mathcal{D}\hat{\varphi}_a e^{-S_{dyn}[\{\hat{\varphi}_a,\varphi_a\}]+\sum_a\int_{xt}(\hat{J}_{a,xt}\varphi_{a,xt}+J_{a,xt}\hat{\varphi}_{a,xt})}
\end{aligned}
\end{equation}
where the (bare) dynamical action is
\begin{equation}
\begin{aligned}
\label{eq_bare_action}
S_{dyn}[\{\hat{\varphi}_a,&\varphi_a\}]= 
\sum_a \int_{xt}\hat{\varphi}_{a,xt}\Big\{\partial_t\varphi_{a,xt}-T\hat{\varphi}_{a,xt}
\\&  +\frac{\delta S_B[\varphi_a]}{\delta{\varphi}_{a,xt}}\Big\} -\frac{\Delta_B}{2}\sum_{ab}\int_{xtt'}  \hat{\varphi}_{a,xt}\hat{\varphi}_{b,xt'}
\end{aligned}
\end{equation}
and where we have set $\Omega_B = 1$; $\ln Z$ is the sought generating functional of the response and correlation functions.

In the long-time limit, the relaxation toward equilibrium satisfies, in addition to the causality requirement, an invariance under time translation (TTI) and a time-reversal symmetry (TRS).\cite{zinnjustin89} The latter in turn implies the fluctuation-dissipation theorem.\cite{janssen92,zinnjustin89,andreanov06} The TRS corresponds to an invariance of the theory under the simultaneous transformations $t \rightarrow -t$, $\varphi_a \rightarrow \varphi_a$, and $\hat\varphi_a \rightarrow \hat\varphi_a-(1/T)\partial_t\varphi_a$.\cite{andreanov06}

\section{nonperturbative functional renormalization group}

The theoretical formalism we use to describe the long-time, long-distance physics of the RFIM near its critical point is the NP-FRG. We have generalized the formalism developed for the (static) equilibrium properties of the RFIM\cite{tarjus04,tissier06,tissier11} by combining it with the approach put forward by Canet et al.\cite{canet}  for the critical dynamics of the Ising model in the absence of quenched disorder.

To  apply the NP-FRG formalism to the above dynamical field theory, we introduce an infrared (IR) regulator $\Delta S_k$ to the action (\ref{eq_bare_action}), whose role is to suppress the integration over slow modes associated with momenta $\vert q \vert \lesssim k$ in the functional integral:\cite{wetterich93,berges02,tarjus04,tissier11}
\begin{equation}
\begin{aligned}
\label{bare_action}
&\Delta S_{k}[\{\Phi_a\}]= \frac 12 \int_{xx'tt'}{\rm{tr}}\Big[\sum_a \Phi_{a,xt}\widehat{\mathbf R}_k(\vert x-x'\vert,t-t') \Phi_{a,x't'}^\top \\&
+\frac 12 \sum_{ab} \Phi_{a,xt}\widetilde{\mathbf R}_k(\vert x-x'\vert,t-t') \Phi_{b,x't'}^\top\Big] \,,
\end{aligned}
\end{equation}
where $\Phi_a\equiv (\varphi_a,\hat\varphi_a)$,  $\Phi_a^\top$ its transpose, and $\widehat{\mathbf R}_k$ and $\widetilde{\mathbf R}_k$ are symmetric $2\times 2$ matrices of mass-like IR cutoff functions that enforce the decoupling between fast (high-momentum) and slow (low-momentum) modes in the partition function. Following Ref. [\onlinecite{canet}] it proves sufficient to control the contribution of the fluctuations through their momentum dependence and take $\widehat R_{k,11}=\widehat R_{k,22}=0$,  $\widehat R_{k,12}=\widehat R_{k,21}=\widehat R_k(x-x')$, and $\widetilde R_{k,11}=\widetilde R_{k,12}=\widetilde R_{k,21}=0$, $\widetilde R_{k,22}=\widetilde R_k(x-x')$ where  $\widehat R_k(q^2)$ and $\widetilde R_k(q^2)$ are chosen (in Fourier space) such that the integration over modes with momentum $\vert q \vert \lesssim k$ is suppressed.\cite{berges02,tarjus04,tissier11} To avoid an explicit breaking of the underlying super-rotations of the theory,\cite{parisi79} we take\cite{tissier11}
\begin{equation}
\label{eq_ward_surot}
\widetilde R_k(q^2)\propto \frac{\partial \widehat R_k(q^2)}{\partial q^2}\, .
\end{equation}
Note that the above choice of IR regulator satisfies the TRS, a crucial property.

Through this addition $Z[\{\mathcal J_a\}]$ is replaced by a $k$-dependent quantity, $Z_k[\{\mathcal J_a\}]$, where $\mathcal J_a$ denotes $(\hat J_a,J_a)$. The  central quantity of the NP-FRG is the ``effective average action" $\Gamma_k$,\cite{wetterich93} which is the generating functional of the 1-particle irreducible correlation functions at the scale $k$. It is defined (modulo the subtraction of a regulator contribution) from $\ln Z_k[\{\mathcal J_a\}]$ via a Legendre transform:
\begin{equation}
\label{eq_legendre_transform_k}
\Gamma_k[\{\Phi_a\}]+\ln{Z_k[\{\mathcal J_a\}]}= \sum_a \int_{xt} {\rm{tr}} \mathcal{J}_{a,xt}\Phi_{a,xt}^\top-\Delta S_k[\{\Phi_a\}],
\end{equation}
where $\Phi_a\equiv (\phi_a,\hat\phi_a)$ now denotes the ``classical'' (or average) fields with $\phi_{a,xt}=\delta \ln Z_k/\delta \hat J_{a,xt}=\langle \varphi_{a,xt}\rangle$ and $\hat{\phi}_{a,xt}=\delta \ln Z_k/\delta J_{a,xt}=\langle \hat\varphi_{a,xt}\rangle$; the trace operation is over the 2 components of $\Phi_a$ and $\mathcal J_a$. 

Expansions in generalized cumulants are then generated by expanding the functionals in increasing number of unrestricted sums over copies,
\begin{equation}
\label{eq_expansion_gamma_k}
\Gamma_k[\{\Phi_a\}]=\sum_{p=1}^\infty \frac {(-1)^{p-1}}{p!}\sum_{a_1\cdots a_p} \mathsf \Gamma_{kp}[\Phi_{a_1},\cdots,\Phi_{a_p}]
\end{equation}
where $\mathsf \Gamma_{kp}$ can be formally expressed as
\begin{equation}
\begin{aligned}
\label{eq_gamma_kp}
\mathsf \Gamma_{kp}=\int_{x_1t_1\cdots x_p t_p}\hat\phi_{a_1,x_1t_1}\cdots \hat\phi_{a_p,x_pt_p} \gamma_{k p;x_1t_1,\cdots,x_pt_p}
\end{aligned}
\end{equation}
with $\gamma_{k p}$ a functional of the fields  $\Phi_{a_1,t_1},\cdots, \Phi_{a_p,t_p}$ and of their time derivatives, $\partial_{t_1}^q \Phi_{a_1,t_1},\cdots, \partial_{t_p}^q \Phi_{a_p,t_p} $, $q\geq 1$. When the fields are chosen uniform in time with $\phi_{a,xt}=\phi_{a,x}$ and $\hat\phi_{a,xt}=0$, the $\gamma_{k p}$'s reduce to the cumulants of the renormalized random field at equilibrium already studied in Refs. [\onlinecite{tarjus04,tissier06,tissier11}]. For generic fields, the additional contributions then represent kinetic terms.\cite{balents-ledoussal}

The functional $\Gamma_k$ satisfies an exact RG equation (ERGE) that describes its evolution with the IR cutoff $k$,\cite{wetterich93}
\begin{equation}
\label{eq_wetterich}
\partial_k\Gamma_k[\{\Phi_a\}]=\frac 12 \rm{Tr} \big\{ (\partial_k\mathbf R_k) (\Gamma_k^{(2)}[\{\Phi_a\}] + \mathbf R_k)^{-1} \big\},
\end{equation}
where the trace is over space-time coordinates, copy indices and components, and $\Gamma_k^{(2)}$ is the the matrix formed by the second functional derivatives of $\Gamma_k$. (In what follows, superscripts within a parenthesis are used to indicate derivatives with respect to the appropriate arguments.) By inserting the expansion in increasing number of sums over copies and proceeding to the associated algebraic manipulations, one then derives an infinite hierarchy of ERGE's for the generalized cumulants $\mathsf \Gamma_{kp}$ or, alternatively, for the functionals $\gamma_{k p}$.

\section{nonperturbative approximation scheme}

One cannot hope to solve the infinite hierarchy of functional flow equations for the $\gamma_{k p}$'s exactly, but one can describe the long-distance physics of the problem by means of a nonperturbative approximation scheme. We combine the minimal truncation of the effective average action already shown to successfully describe the equilibrium critical behavior of the RFIM\cite{tissier11} with an account of the dynamics through a truncation of the expansion in kinetic coefficients that allows us to describe the characteristic relaxation time. By taking into account the TRS, we arrive at the following ansatz:
\begin{equation}
\begin{aligned}
\label{eq_gammak1_ansatz}
\gamma_{k1;xt}[\Phi]=&\frac{\delta}{\phi_{xt}}\Big [U_k(\phi_{xt})+ \frac{1}{2}Z_k(\phi_{xt})(\partial_x\phi_{xt})^2 \Big ] \\& + X_k(\phi_{xt})(\partial_t\phi_{xt}-T \hat{\phi}_{xt})
\end{aligned}
\end{equation}
\begin{equation}
\begin{aligned}
\label{eq_gammak2_ansatz}
\gamma_{k2;x_1t_1,x_2t_2}[\Phi_1,\Phi_2]=\delta^{(d)}(x_1-x_2)\Delta_k(\phi_{1,x_1t_1},\phi_{2,x_1t_2})
\end{aligned}
\end{equation} 
while the $\gamma_{k p}$'s with $p \geq 3$ are set to zero. Note that this ansatz describes the characteristic relaxation time but not its distribution: to do this the next orders of the truncation of the expansion in kinetic coefficients would be required, as discussed in Refs. [\onlinecite{balents-ledoussal,gorokhov_creep}].

The next step is to derive the RG flow equations for the functions contained in the ansatz from the ERGE's for the $\gamma_{k p}$'s. As already mentioned, we work in the It\={o} discretization scheme and the corresponding prescription can be systematically implemented in the NP-FRG equations by following the simple procedure developed in Ref. [\onlinecite{canet}].  The derivation is tedious but straightforward and more details are given below and in the appendix. The output is a set of coupled  flow equations for 3 functions of one field, $U'_k(\phi)$, $Z_k(\phi)$, $X_k(\phi)$,  and one function of 2 fields $\Delta_k(\phi_1,\phi_2)$. As a result of the TRS, the renormalized kinetic function $X_k$ does not enter the flow of the static ones ($U'_k$, $Z_k$ and $\Delta_k$).

The flow equations involve the renormalized propagators at scale $k$ evaluated at the lowest order of the expansions is sums over copies and for fields that are uniform in space and time, $\Phi_{a,xt}\equiv (\phi_a,0)$. In Fourier (momentum) space, these propagators are expressed as $\mathbf P_{k,ab}(q;t,t')=\widehat{\mathbf P}_k(q;\phi_a;t,t')\delta_{ab}+\widetilde{\mathbf P}_{k}(q;\phi_a,\phi_b;t,t')$ where the $2 \times 2$ matrix $\widehat{\mathbf P}_k$ has a structure following from causality, TTI and fluctuation-dissipation theorem, with
\begin{equation}
\label{eq_hat_prop12}
\widehat P_k^{12}(q;\phi;t'-t)=\Theta(t'-t)X_k(\phi)^{-1}e^{-\frac{(t'-t)}{\tau_k(q;\phi)}}
\end{equation}
the response function, $\widehat P_k^{21}(t'-t)=\widehat P_k^{12}(t-t')$, $\widehat P_k^{11}$ the 2-time disorder-connected correlation function given by
\begin{equation}
\label{eq_hat_prop11}
\widehat P_k^{11}(q;\phi;t'-t)=T[Z_k(\phi)q^2+\widehat R_k(q^2)+U''_k(\phi)]^{-1}e^{-\frac{\vert t'-t \vert}{\tau_k(q;\phi)}}
\end{equation}
and $\widehat P_k^{22}=0$; the characteristic relaxation time is defined as
\begin{equation}
\label{eq_tau}
\tau_k(q;\phi)= \frac{X_k(\phi)}{[Z_k(\phi)q^2+\widehat R_k(q^2)+U''_k(\phi)]}\,. 
\end{equation}
In addition, the only nonzero component of $\widetilde{\mathbf P}_{k}$ is 
\begin{equation}
\begin{aligned}
\label{eq_tilde_prop}
&\widetilde P_k^{11}(q;\phi_1,\phi_2;t'-t)= \\&\int_t\widehat P_k^{12}(q;\phi_1;t)\int_{t'}\widehat P_k^{21}(q;\phi_2;t')[\Delta_k(\phi_1,\phi_2)-\widetilde R_k(q^2)]
\end{aligned}
\end{equation}
which, in this truncation, is simply the static (equilibrium) disorder-disconnected correlation function.

Finally, to study the vicinity of the relevant zero-temperature critical fixed point\cite{villain-fisher} the NP-FRG equations are cast in a dimensionless form  by introducing appropriate scaling dimensions: $\phi \sim k^{(d-4+\bar \eta)/2}$, $Z_{k} \sim k^{-\eta}$, $U'_k \sim k^{(d-2\eta+\bar \eta)/2}$, $\Delta_k \sim k^{-(2\eta- \bar \eta)}$ and the renormalized temperature $T_k \sim T k^{\theta}$, where $\theta=2+\eta-\overline{\eta}>0$.\cite{tarjus04,tissier11} We express the results in terms of the dimensionless fields $\varphi=\frac{\varphi_1+\varphi_2}{2}$ and $\delta\varphi=\frac{\varphi_2-\varphi_1}{2}$. With lowercase letters denoting  dimensionless quantities, one can formally write the flow equations for the static quantities as
\begin{equation}
\begin{aligned}
\label{eq_flow_static}
&k \partial_k u'_k(\varphi)=\beta_{u'0}(\varphi)+T_k\beta_{u'1}(\varphi)\,,\\&
k \partial_k z_k(\varphi)=\beta_{z0}(\varphi)+T_k\beta_{z1}(\varphi)\,,\\&
k \partial_k \delta_k(\varphi,\delta\varphi)=\beta_{\delta0}(\varphi,\delta\varphi)+T_k\beta_{\delta1}(\varphi,\delta\varphi)\,,
\end{aligned}
\end{equation} 
where the beta functions depend on $u'_k$, $z_k$, $\delta_k$, their derivatives, and on the (dimensionless) cutoff functions. 

These equations generalize to nonzero temperature those given in Ref. [\onlinecite{tissier11}]: $\beta_{u'0}(\varphi)$, $\beta_{z0}(\varphi)$, and $\beta_{\delta0}(\varphi, \delta\varphi)$ coincide with the zero-temperature beta functions explicitly given in this reference. The beta functions $\beta_{u'1}(\varphi)$ and $\beta_{z1}(\varphi)$ are regular functions whose expression is unilluminating and is not given here. Finally, $\beta_{\delta1}(\varphi, \delta\varphi)$ is given by
\begin{equation}
\begin{aligned}
\label{eq_beta_delta1}
&\beta_{\delta1}(\varphi,\delta\varphi)=
-\frac 18 \int_{\hat q} \widehat\partial_s \hat r(\hat q^2) \Big \{[\widehat p_k(\hat q;\varphi+\delta\varphi)^2+ \rm{sym}] \, \times \\&
[\delta^{(02)}_k(\varphi,\delta\varphi)+\delta^{(20)}_k(\varphi,\delta\varphi)] +2 [\widehat p_k(\hat q;\varphi+\delta\varphi)^2-\rm{sym}]\, \times \\&\delta^{(11)}_k(\varphi,\delta\varphi) + 2\delta^{(01)}_k(\varphi,\delta\varphi) \frac{\partial}{\partial \delta\varphi}[\widehat p_k(\hat q;\varphi+\delta\varphi)^2 +\rm{sym}] \\&
+ 2\delta^{(10)}_k(\varphi,\delta\varphi) \frac{\partial}{\partial \delta\varphi}[\widehat p_k(\hat q;\varphi+\delta\varphi)^2 -\rm{sym}]
\end{aligned}
\end{equation} 
where  $\hat q=q/k$, $\int_{\hat q}\equiv \int d^d \hat q/(2\pi)^d$,  the dimensionless cutoff function $\hat r(\hat q^2)$ is defined through $\widehat R_k(q^2)=Z_k q^2\hat r(\hat q^2)$ and $\widehat\partial_s \hat r(\hat q^2)\equiv -[\eta \hat{q}^2 \hat r(\hat{q}^2)+2 \hat{q}^4 \hat r'(\hat{q}^2)]$ is a symbolic notation for the term obtained from $k\partial_k \widehat R_k(q^2)$. Similarly, one defines $\tilde r(\hat q^2)$ from $\widetilde R_k(q^2)=\Delta_k \tilde r(\hat q^2)$ but it is simply related to $\hat r$ via $\tilde r(\hat q^2)=-\partial_{\hat q^2}[\hat q^2 \hat r(\hat q^2)]$ from Eq. (\ref{eq_ward_surot}). The dimensionless hat propagator is given by $\widehat p_k(\hat q;\varphi)=(\hat q^2[z_k(\varphi)+\hat r(\hat q^2)]+u''_k(\varphi))^{-1}$. Finally, $\rm{sym}$ denotes a term obtained by changing $\delta\varphi$ in $-\delta\varphi$.

When $T=0$, it was previously found that the fixed-point solution displays two regimes:\cite{tissier11} 

(1) for $d <d_{DR}\simeq 5.1$, a ``cusp" in $\vert \delta\varphi \vert$ is present in the fixed-point function $\delta_*$ when $\delta\varphi\rightarrow 0$:
\begin{equation}
\begin{aligned}
\label{eq_cusp}
\delta_*(\varphi,\delta\varphi)=\delta_{*}(\varphi,0)-\delta_{*,a}(\varphi)\vert \delta\varphi \vert +O(\delta\varphi^2)
\end{aligned}
\end{equation} 
with $\delta_{*,a}\neq 0$. This cusp is associated with the presence of avalanches on all scales at the critical point.\cite{tarjus13}

(2) For $d>d_{DR}$ the fixed-point function $\delta_*$ is ``cuspless", which ensures that the $d\rightarrow d-2$ dimensional-reduction property of the (static) critical exponents\cite{parisi79} is valid (and that the super-rotation is not spontaneously broken along the RG flow). Avalanches are still present but lead to only a subdominant cusp,\cite{tarjus13}
\begin{equation}
\begin{aligned}
\label{eq_subdominant_cusp}
\delta_k(\varphi,\delta\varphi)=\delta_{*}(\varphi,0)-\delta_{k,a}(\varphi)\vert \delta\varphi \vert +O(\delta\varphi^2)
\end{aligned}
\end{equation} 
when $k\rightarrow 0$, where $\delta_{k,a}$ goes to zero as
\begin{equation}
\begin{aligned}
\label{eq_scaling_cusp}
\delta_{k,a}(\varphi)\sim k^\lambda
\end{aligned}
\end{equation} 
with $\lambda >0$ characterizing the (diverging) number of spanning avalanches.\cite{dahmen96,tarjusNEW}

\section{thermal boundary layer}

Describing  the critical slowing down requires a nonzero temperature and additional care is needed. The beta function for $\delta_k$  shows a nonuniform convergence when $k\rightarrow 0$ and $\delta\varphi\rightarrow 0$ and for a nonzero $T$ the cusp is rounded near $\delta\varphi=0$ in a ``boundary layer'' of width $\delta\varphi\sim T_k$ (see also Refs.~[\onlinecite{chauve_creep,balents-ledoussal}] for the case of the elastic manifold in a random environment). 

For $T>0$ the beta function of $\delta_k$ in the limit $\delta\varphi\to 0$ and $T_k\rightarrow 0$  can indeed be written as
\begin{equation}
\begin{aligned}
\label{eq_beta_delta_asympt}
k \partial_k \delta_k(\varphi,\delta\varphi) \simeq& \beta_{\delta,reg}(\varphi)+
\frac{a_{1k}(\varphi)}{2}\frac{\partial^2[\delta_k(\varphi,\delta\varphi)-\delta_k(\varphi,0)]^2}{\partial(\delta\varphi)^2}
 \\&- T_k a_{2k}(\varphi)\delta^{(02)}_k(\varphi,\delta\varphi)
\end{aligned}
\end{equation} 
where $\beta_{\delta,reg}$ is the contribution that is independent of the derivatives of $\delta_k$ with respect to $\delta\varphi$; $a_{1k}$ and $a_{2k}$ are regular functions obtained from the static functions: $a_{1k}(\varphi)$ is the prefactor of the anomalous contribution in $\beta_{\delta0}(\varphi,\delta\varphi)$ whose limit when $\delta\varphi \rightarrow 0$ is nonzero only in the presence of a cusp,
\begin{equation}
\begin{aligned}
\label{eq_a1k}
a_{1k}(\varphi)= \frac 12 \int_{\hat q} \partial_s \hat r(\hat q^2) \widehat p_k(\hat q;\varphi)^3 \, ,
\end{aligned}
\end{equation} 
and $a_{2k}(\varphi)$ is the prefactor of the potentially singular piece in $\beta_{\delta1}(\varphi,\delta\varphi)$,
\begin{equation}
\begin{aligned}
\label{eq_a2k}
a_{2k}(\varphi)=\frac 14 \int_{\hat q} \partial_s \hat r(\hat q^2) \widehat p_k(\hat q;\varphi)^2 \,.
\end{aligned}
\end{equation} 

The cusp in $\vert \delta\varphi \vert$ that is present in $T=0$ is rounded at finite temperature because of the last term in Eq. (\ref{eq_beta_delta_asympt}). Instead, $\delta_k$ develops a ``thermal boundary layer'',
\begin{equation}
\label{eq_BL_ansatz}
 \delta_k(\varphi,\delta\varphi)=\delta_{k}(\varphi,0)+T_k b_k(\varphi,y=\frac{\delta\varphi}{T_k})+O(T_k^2,\delta\varphi^2)
\end{equation}
when $T_k,\delta\varphi \rightarrow 0$. It is easy to derive that the solution has the explicit form 
\begin{equation}
\label{eq_BL_solution}
b_k(\varphi,y)=\frac{a_{2*}(\varphi)}{a_{1*}(\varphi)}\Big(1-\sqrt{1+\frac{a_{1*}(\varphi)^2\delta_{k,a}(\varphi)^2}{a_{2*}(\varphi)^2}y^2}\Big),
\end{equation}
where $a_{p*}$ are the (nonzero) fixed-point functions and $\delta_{k,a}(\varphi)$ behaves differently when $k\rightarrow 0$ for $d <d_{DR}$ and $d> d_{DR}$ (see above).

\section{activated dynamic scaling}

We now turn to the NP-FRG equation for the kinetic term $X_k(\phi)$. It is obtained from the renormalization prescription
\begin{equation}
\begin{aligned}
\label{eq_RGprescr_X_k}
X_k(\phi)=-\frac 1T \frac{\partial}{\partial \hat\phi}\gamma_{k1;xt}(\phi,\hat\phi)\Big\vert_{\hat\phi=0}
\end{aligned}
\end{equation} 
and is given in graphical terms in the appendix. After having introduced the dimensionless quantities, it can be rewritten as
\begin{equation}
\begin{aligned}
\label{eq_FRG_Xk_expl1}
k\partial_k X_k(\varphi)= \beta_{X0}(\varphi)+T_k \beta_{X1}(\varphi)
\end{aligned}
\end{equation} 
where $\beta_{X0}(\varphi)$ is given in the appendix and $\beta_{X1}(\varphi)$ is a regular function that leads only to subdominant terms near the fixed point. 

Note that we have kept $X_k$ itself in a dimensionful form. In the case of a conventional critical slowing down,\cite{hohenberg-halperin} one introduces a dynamical exponent $z$ such that the characteristic relaxation time scales as $\tau_k\sim k^{-z}$ near the fixed point. The kinetic term then has dimension $X_k \sim k^{-z+(2-\eta)}$.\cite{canet}  However, in the presence of a nonzero random-field strength, where one anticipates an unconventional activated dynamic scaling, one should rather focus on $F_k=\ln X_k$ (where if needed $X_k$ can be made dimensionless inside the logarithm by dividing by a $k$-independent factor).

By inserting the boundary layer solution in Eqs.~(\ref{eq_FRG_Xk_expl1}) and (\ref{eq_beta_Xk0_app}) (see the appendix) and working at the dominant orders when $T_k\rightarrow 0$ (and $k\rightarrow 0$), it is easy to derive the flow of $F_k(\varphi)=\ln X_k(\phi)$, which reads
\begin{equation}
\begin{aligned}
\label{eq_flow_Fk}
& k \partial_k F_k(\varphi)=\frac{d-4+\overline{\eta}}{2}\varphi F_k'(\varphi)+\tilde c_{1k}(\varphi)+\tilde c_{2k}(\varphi)F_k'(\varphi)  + \\&
\tilde c_{3k}(\varphi)[F_k''(\varphi)+F_k'(\varphi)^2]-\frac{1}{T_k}\frac{a_{1,k}(\varphi)^2}{a_{2k}(\varphi)} \delta_{k,a}(\varphi)^2+O(T_k)\,,
\end{aligned}
\end{equation} 
where the $\tilde c_{p k}$'s are regular functions of $\varphi$ whose expressions are given in the appendix.

The solution of Eq.~(\ref{eq_flow_Fk}) when $k\rightarrow 0$ is then of the form 
\begin{equation}
\label{eq_Fk}
F_k(\varphi)=\frac{1}{T_k}e_k + \frac{1}{\sqrt{T_k}}g_k(\varphi)+O(1)
\end{equation}  
with $e_k$ independent of $\varphi$. We find that $e_k\rightarrow e_*>0$ for $d <d_{DR}$ and $e_k\rightarrow 0$ as $k^{2\lambda}$ for $d >d_{DR}$ [compare with the term in $1/T_k$ in Eq. (\ref{eq_flow_Fk})].

\section{results}

To support the analytical solutions provided above and compute the exponents we have numerically solved the NP-FRG equations for a wide range of dimensions between 3 and 6. To do so, we have discretized the fields on a grid and used a variation of the Newton-Raphson method. For the cutoff function $\hat r_k(\hat q^2)$ we have chosen the same form as in previous work and optimized the parameters by stability considerations.\cite{tissier11}

\begin{figure}[h!]
\begin{center}
\includegraphics[width=200pt,height=133pt]{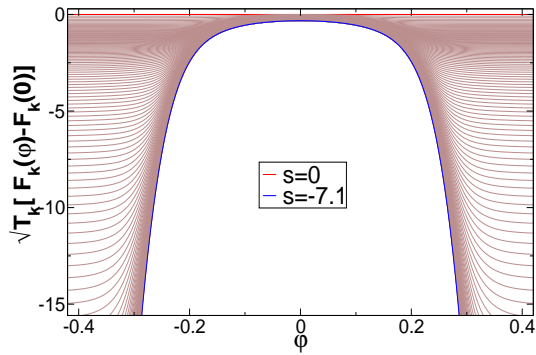}  
\caption{Evolution of $\sqrt{T_k} [F_k(\varphi)-F_k(\varphi_0)]$ for RG times $s=\ln k$ from $0$ (initial condition where the function is chosen equal to zero: red line) to -7.1 (essentially, the fixed point: blue line). Here $\varphi_0=0$ (but other choices lead to the  same asymptotic function). The function is illustrated for the case $d=4.4$ and a bare temperature $T=0.1$; the minimum of the effective potential is then for $\varphi \simeq \pm 0.085$.}
\label{figS1}
\end{center}
\end{figure}

First, the numerical solution confirms the behavior of  $F_k(\varphi)$ given in Eq. (\ref{eq_Fk}). We illustrate this point by showing the flow of  $\sqrt{T_k} [F_k(\varphi)-F_k(0)]$, which should asymptotically converge to $g_k(\varphi)-g_k(0)$, for the case $d=4.4$. We can see from Fig. \ref{figS1} that the fixed-point function is indeed well-behaved.

\begin{figure}[h!]
\begin{center}
\includegraphics[width=200pt,height=133pt]{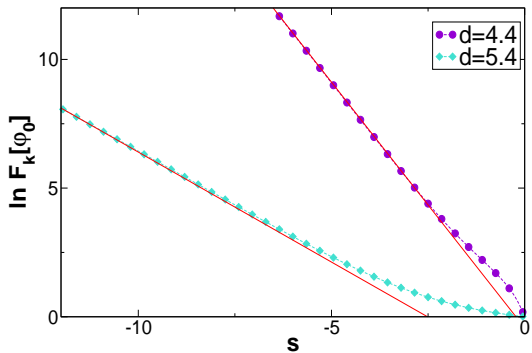}  
\caption{The flow of $\ln F_k(\varphi_0)$ with the RG ``time'' $s=\ln(k/\Lambda)$ with the UV scale $\Lambda \equiv 1$; $\varphi_0$ is (arbitrarily) chosen near the minimum of $u_k(\varphi)$, for $d=5.4$ and $d=4.4$. The thin full lines denote the expected asymptotic slopes, $\psi=\theta\approx 1.909$ for $d=4.4$ and $\psi=\theta-2\lambda\approx 0.855$ for $d=5.4$.}
\label{fig2}
\end{center}
\end{figure}

In Fig.~\ref{fig2} we further illustrate the dominant $1/T_k$ dependence of $F_k$ for the two cases discussed above with $F_k\sim 1/T_k$ for $d=4.4 <d_{DR}$ and $F_k\sim k^{2\lambda}/T_k$ for $d=5.4>d_{DR}$.

As a result of Eqs. (\ref{eq_tau}) and (\ref{eq_Fk}), the asymptotic behavior of $\ln \tau_k$ goes as $e_k/T_k$ with $e_k$ given above and $T_k\sim T k^\theta$, so that the characteristic relaxation time $\tau_k$  follows the activated dynamic scaling with 
\begin{equation}
\begin{aligned}
\label{eq_tau_k}
\tau_k \sim e^{\frac{E}{T}k^{-\psi} }
\end{aligned}
\end{equation} 
for its leading behavior when $k\rightarrow 0$ (we have dropped the prefactor and subdominant terms in the exponential). 

\begin{figure}
\begin{center}
\includegraphics[width=200pt,height=133pt]{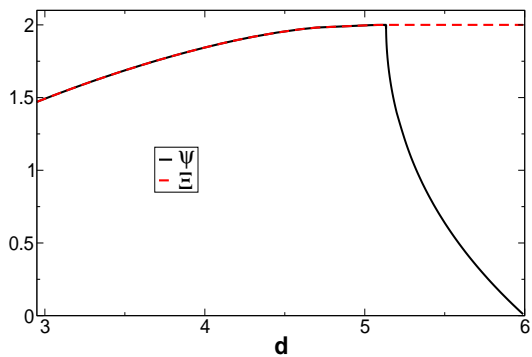}  
\caption{The barrier exponent $\psi$ (full black line) and the temperature exponent $\theta$ (dashed red line) versus dimension $d$; $\theta \simeq 1.49$ in $d=3$ and $1.84$ in $d=4$, in excellent agreement with simulation results.\cite{simulation}}
\label{fig3}
\end{center}
\end{figure}

From the above, we find $\psi=\theta$ for $d<d_{DR}$ and $\psi=\theta- 2\lambda$ with $\theta=2$ (due to dimensional reduction) and $\lambda>0$ for $d>d_{DR}$. At the upper critical dimension $d=6$, one finds that $\lambda=1$ around the Gaussian fixed point so that $\psi=0$.\cite{footnoteLOG} Activated dynamic scaling thus gives way to conventional dynamic scaling,\cite{footnoteRM} a scaling which is for instance found in the mean-field, fully connected, version of the model. The barrier exponent $\psi$ is shown as a function of $d$ in Fig. \ref{fig3}.

\section{discussion}

The above results can be physically interpreted by invoking the collective events that are present at criticality. At $T=0$ these are avalanches, which have a fractal dimension $d_f=(d+4-\bar\eta)/2-2\lambda$ with $\lambda=0$ for $d<d_{DR}$ and $\lambda>0$ for $d>d_{DR}$.\cite{tarjus13} For $T>0$ but very small, there emerge from these avalanches, which correspond to an exceptional degeneracy  between ground states, low-energy excitations corresponding to \textit{quasi}-degeneracy, known as droplets. The result for the barrier exponent can then be rationalized by assuming that with a probability $T/L^{(\theta-2\lambda)}$ there are critical samples of size $L$ with such a quasi-degeneracy and that the whole energy landscape of the quasi-degenerate system has a unique scale, then given by $L^{(\theta-2\lambda)}$. 

The droplet picture also predicts anomalous static thermal fluctuations of the field (or magnetization).\cite{balents-ledoussal-droplets} In the rare samples, the magnitude of the magnetization fluctuations goes as $L^{d_f}$, so that the $p$th cumulants associated with the thermal fluctuations of the magnetization, $\overline{[L^{-d}<(\int_x\varphi_x-<\int_x\varphi_x>)^2>]^p}$, has an anomalous scaling, $\propto (T/L^{\theta-2\lambda}) L^{(4-\bar\eta-2\lambda)p}$. We have checked the validity of the scaling for $p=2$ from the NP-FRG equations, along the lines detailed in Ref. [\onlinecite{tissier06}].

Finally, we conclude by proposing a way to directly check the two regimes, $\psi=\theta$ and $\psi=\theta-2\lambda$ with $\lambda>0$, in computer simulations. As recently shown,\cite{balog} the 1-$d$ RFIM with long-range power-law interactions $\propto \vert x\vert^{d+\sigma}$ has a critical value $\sigma_{c}\simeq 0.379$ around which the change of regime should be observed. In the mean-field region, for $\sigma<1/3$, one should also find that the relaxation is no longer activated but follows conventional scaling with the mean-field dynamic exponent $z=\sigma$.

\appendix
\section{NP-FRG flow for the kinetic term}

The flow of the kinetic term $X_k(\phi)$ is obtained from the renormalization prescription in Eq. (\ref{eq_RGprescr_X_k}) and from the ERGE for $\gamma_{k1;xt}$. In graphical terms the flow equation reads
\begin{eqnarray}
\label{eq_FRG_Xk_diag}
&&\partial_sX_{k}(\phi)=\frac{1}{4}\tilde{\partial_s}\int_q
\nonumber\\
&&\Bigg(\raisebox{-15pt}{\includegraphics[width=240pt,height=35pt]{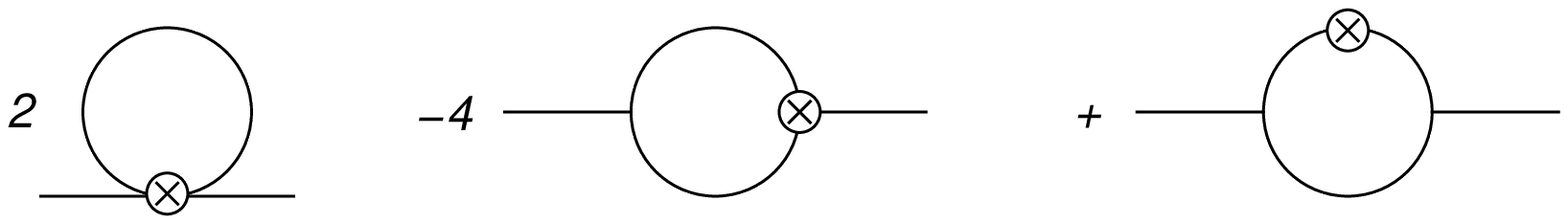}}
\nonumber\\
&&\raisebox{-15pt}{\includegraphics[width=240pt,height=30pt]{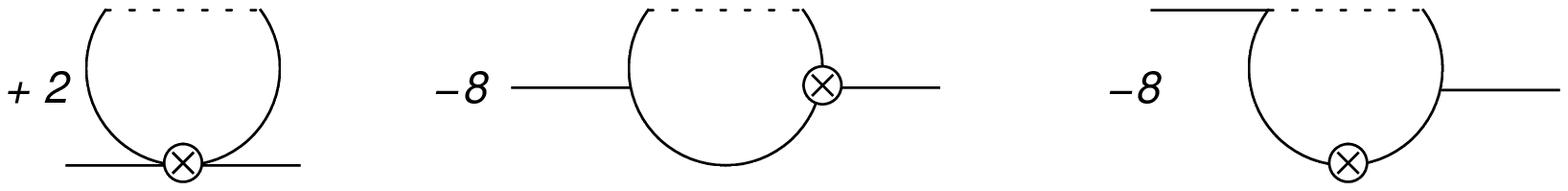}}
\nonumber\\
&&\raisebox{-15pt}{\includegraphics[width=240pt,height=30pt]{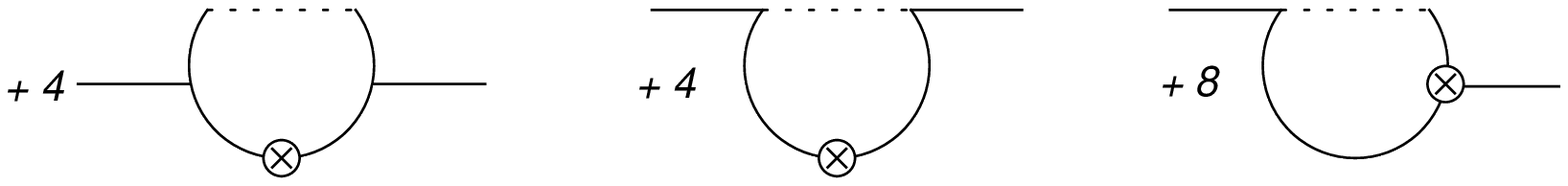}}\Bigg)\nonumber\\
\end{eqnarray}
where a cross in the circle denotes a vertexÓ $X_k(\phi)$, the lines denote static propagators $\widehat P_k$ and the dotted lines the disorder vertex $\Delta_k(\phi_1,\phi_2)$ (after having taken the needed derivatives, one sets $\phi_1=\phi_2=\phi$); $\widetilde \partial_k$ is a short-hand notation to indicate a derivative acting only on the cutoff functions, \textit{i.e.},  $\widetilde \partial_k \equiv \partial_k \widehat R_k \delta/\delta \widehat R_k+\partial_k \widetilde R_k \delta/\delta \widetilde R_k$. For implementing It\={o} prescription we have followed the trick devised by Canet et al.\cite{canet}, which amounts to shifting the time dependence of the response field by an infinitesimal amount in the renormalized response functions. 

With the help of the dimensionless quantities, Eq. (\ref{eq_FRG_Xk_diag}) can be rewritten as
\begin{equation}
\begin{aligned}
\label{eq_FRG_Xk_expl1_appendix}
k\partial_k X_k(\varphi)= \beta_{X0}(\varphi)+T_k \beta_{X1}(\varphi)
\end{aligned}
\end{equation} 
where
\begin{equation}
\begin{aligned}
\label{eq_beta_Xk0_app}
&\beta_{X0}(\varphi)=
\frac{d-4+\overline{\eta}}{2}\varphi X_k'(\varphi) 
+ \frac{1}{4}\int_{\hat{q}}\Bigg\{ -2\widehat\partial_s\tilde{r}(\hat{q}^2) \hat{p}_k(\hat{q};\varphi)^2 \\&
\times X_k''(\varphi) 
- 4 \hat{p}_k(\hat{q};\varphi)^3\big [\hat{q}^2 z_k'(\varphi) + u_k^{(3)}(\varphi) \big ]
\Big(-2 \big [\widehat\partial_s\tilde{r}(\hat{q}^2) \\&
+ 3\widehat\partial_s\hat{r}(\hat{q}^2) \hat{p}_k(\hat{q};\varphi)(\delta_k(\varphi,0)-\tilde{r}(\hat{q}^2) )\big ] X_k'(\varphi) 
- 3\widehat\partial_s\hat{r}(\hat{q}^2) \\&
\times \hat{p}_k(\hat{q};\varphi)\delta_k^{(1, 0)}(\varphi,0) X_k(\varphi) \Big) 
-4 \hat{p}_k(\hat{q};\varphi)^4  X_k(\varphi)\big [\widehat\partial_s\tilde{r}(\hat{q}^2)\\&
+ 4\widehat\partial_s\hat{r}(\hat{q}^2) (\delta_k(\varphi,0)-\tilde{r}(\hat{q}^2)) \hat{p}_k(\hat{q};\varphi) \big ]\big [\hat{q}^2 z_k'(\varphi) + u_k^{(3)}(\varphi) \big ]^2\\&
- 2\widehat\partial_s\hat{r}(\hat{q}^2)\hat{p}_k(\hat{q};\varphi)^3\Big[2 (\delta(\varphi,0)-\tilde{r}(\hat{q}^2)) X_k''(\varphi) 
+ 4\delta_k^{(1, 0)}(\varphi,0) \\&  X_k'(\varphi)
 + \big(-\delta_k^{(0, 2)}(\varphi,0) + \delta_k^{(2, 0)}(\varphi, 0) \big)X_k(\varphi) \Big ] \Bigg\}
\end{aligned}
\end{equation} 
where $\widehat\partial_s \tilde r(\hat q^2)$ is a symbolic notation for $[2 \eta-\bar{\eta}][\hat r(\hat{q}^2)+\hat{q}^2 \hat r'(\hat{q}^2)]
+2[2 \hat{q}^2 \hat r'(\hat{q}^2)+\hat{q}^4 \hat r''(\hat{q}^2)]$. The regular term $\beta_{X1}(\varphi)$ leads only to subdominant terms near the fixed point and is not given here.

The flow of $F_k(\varphi)=\ln X_k(\phi)$ at dominant orders when $T_k\rightarrow 0$ is given in Eq. (\ref{eq_flow_Fk}) and the (regular) functions $\tilde c_{p k}(\varphi)$ are expressed at the relevant order in $T_k$ as
\begin{equation}
\begin{aligned}
\label{eq_prefactors_c1k}
& \tilde{c}_{1k}(\varphi)=\frac{1}{2}\int_{\hat{q}}\Big[-2\widehat\partial_s\tilde{r}(\hat{q}^2) \hat{p}_k(\hat{q};\varphi)^4[\hat{q}^2 z_k'(\varphi) + u_k^{(3)}(\varphi) ]^2 \\&
+ \widehat\partial_s\hat{r}(\hat{q}^2)\hat{p}_k(\hat{q};\varphi)^3\Big(-\delta_k^{(20)}(\varphi,0) + 2 \hat{p}_k(\hat{q};\varphi)[\hat{q}^2 z_k'(\varphi) \\&
 + u_k^{(3)}(\varphi) ]\big [3\delta_k^{(10)}(\varphi,0) - 4 \hat{p}_k(\hat{q};\varphi)(\delta_k(\varphi,0)-\tilde{r}(\hat{q}^2) )\\&
\times  (\hat{q}^2 z_k'(\varphi) + u_k^{(3)}(\varphi) ) \big ] \Big) \Big] +O(T_k) \,,
\end{aligned}
\end{equation} 
\begin{equation}
\begin{aligned}
\label{eq_prefactors_c2k}
&\tilde{c}_{2k}(\varphi)=2\int_{\hat{q}} \hat{p}_k(\hat{q};\varphi)^3 \Big[\widehat\partial_s\tilde{r}(\hat{q}^2)(\hat{q}^2 z_k'(\varphi) + u_k^{(3)}(\varphi) )\\&
 + \widehat\partial_s\hat{r}(\hat{q}^2)\Big(-\delta_k^{(10)}(\varphi,0)
+ 3 \hat{p}_k(\hat{q};\varphi)[\delta_k(\varphi,0) -\tilde{r}(\hat{q}^2) ]\\&
\times [\hat{q}^2 z_k'(\varphi) + u_k^{(3)}(\varphi) ] \Big) \Big] +O(T_k) \,,
\end{aligned}
\end{equation} 
\begin{equation}
\begin{aligned}
\label{eq_prefactors_c3k}
&\tilde c_{3k}= - \frac 12 \int_{\hat q} \Big (\widehat\partial_s \tilde r(\hat q^2)+2\widehat\partial_s \hat r(\hat q^2)[\delta_k(\varphi,0)-
\tilde r(\hat q^2)] \widehat p_k(\hat q;\varphi)\Big ) \\& \times
\widehat p_k(\hat q;\varphi)^2 -T_k\int_{\hat q} \widehat\partial_s \hat r(\hat q^2)\widehat p_k(\hat q;\varphi)^2\,.
\end{aligned}
\end{equation}

\begin{acknowledgments}
We acknowledge support from G. Biroli's ERC grant NPRGGLASS.  
\end{acknowledgments}


\begin{thebibliography}{15}
\bibitem{footnote_defDR}
The dimensional-reduction property means that the critical exponents of the RFIM in dimension $d$ are identical to those of the pure Ising model in dimension $d-2$.
\bibitem{footnote_DR}
Note that dimensional reduction does not apply to the dynamics even in perturbation theory. The exponent $z$ found in Ref. [\onlinecite{krey84}] differs from that of the pure Ising model in two dimensions less. The invariance under the supersymmetry (super-rotation) that leads to the $d\rightarrow d-2$ dimensional reduction for the critical exponents describing the main scaling of the static quantities is never satisfied by the kinetic piece of the dynamical theory.
\bibitem{footnote_replicas}
One main advantage of introducing copies or replicas is that the cumulants of the renormalized disorder can be generated by expansions in increasing number of unrestricted sums over copies (or free replica sums). One could try to avoid replicas in the dynamical formalism by replacing the expansion in free replica sums by an expansion in increasing number of integrals over time. However, disentangling this nonlocal time-dependence from the quasi-local dependence associated with time derivatives is an intricate problem.
\bibitem{footnoteRM}
This should be contrasted with the behavior of pinned elastic manifolds  for which dynamics is activated even at the upper critical dimension $d=4$ and $\psi=\theta$.\cite{balents-ledoussal}
\end{thebibliography}
\end{document}